\documentclass[aps,prl,preprint,groupedaddress]{revtex4}
\usepackage{graphicx}
\usepackage{dcolumn}
\begin{document}

\title{Bias in the temperature of helium nanodroplets measured by
an embedded rotor}

\author{Kevin K. Lehmann}
\email[]{Lehmann@princeton.edu}
\affiliation{Department of Chemistry, Princeton University, 
Princeton NJ 08544}

\date{\today}

\begin{abstract}
The ro--vibrational spectra of molecules dissolved in liquid $^4$He
nanodroplets display rotational structure.
Where resolved, this structure has been used to determine a temperature
that has been assumed to equal that of the intrinsic excitations of
the helium droplets containing the molecules.  Consideration of the density
of states as a function of energy and total angular momentum
demonstrates that there is a small but significant bias of the rotor populations
that make the temperature extracted from a fit to its rotational
level populations slightly higher than the temperature of the ripplons of
the droplet.   This bias grows with both the total angular momentum
of the droplet and with the moment of inertia of the solute molecule.
\end{abstract}


\maketitle

\section{Introduction}

A hallmark of ro--vibrational spectroscopy of molecules in liquid $^4$He nanodroplets
has been the observation of rotational structure, as in the gas
phase~\cite{Hartmann95,Callegari01}.    The distribution of intensity of transitions in
this structure has allowed the determination of the rotational temperature of these
molecules, and it has been assumed that this provides a measure of the temperature of the
droplets themselves~\cite{Hartmann95,Grebenev98}.  This assumption has been supported by
the close similarity of the temperatures  determined for a wide range of molecules and by
the assumption that on the long time scales between pickup of solutes and spectroscopic
interrogation, the entire system will come to equilibrium as the droplets cool by
evaporation.

Adriaan Dockter and I have modeled the evaporative cooling of pure and
doped helium nanodroplets using an angular momentum conserving statistical rate
model~\cite{Dokter03}.   These calculations have predict that the droplets cool to a
temperature close to those inferred from experiments and previous evaporative cooling
calculations, but with a vastly broader distribution of energy and total angular momentum
than for a canonical distribution at the same temperature.   
We also found, much to our initial surprise, that while the populations of the
rotational levels of a solvated rotor follow a thermal distribution, the
fitted temperature of the rotor populations is higher than that of the droplet itself.  
While this would seem to violate a fundamental principle of thermodynamics,
as will be demonstrated below, this divergence of the two temperatures 
is in fact a consequence of the constraints imposed by conservation of
angular momentum.   The temperature of the helium is defined by the inverse of
the derivative of the log of the  density of states with respect to energy at fixed total
angular momentum.   However, when a rotor is excited to a state with rotor angular momentum
quantum number $j$, the internal degrees of freedom of the droplet can have angular
momentum values between $|J-j|$ and $J+j$, where $J$ is the total angular momentum quantum
number, which is treated as a conserved quantity.  Thus the derivatives of the density of
droplet states with respect to $J$ contributes to the rotor population.  

\section{State Distributions}
At the temperature of helium nanodroplets, $~0.38\,$K, the only intrinsic 
droplet excitations are surface ripplon modes~\cite{Brink90}.  The density of states and
other thermodynamic quantities can be written as a function of a reduced energy,
$\bar{E} = E / E_{\rm R}$ where $E_{\rm R} = 3.77\, k_{\textrm{B}} \textrm{K}/\sqrt{N}$
and $N$ is the number of helium atoms in the droplet~\cite{Brink90}.  We denote the
density of states as a function of reduced energy and total angular
momentum as $\rho_{{\rm R}J}(\bar{E},J)$, and the density states summed over
angular momentum states as $\rho_{\rm R}(\bar{E}) = 
\sum_J (2J+1) \rho_{{\rm R}J}(\bar{E},J)$.  Using methods
recently described~\cite{Lehmann03}, these can be well approximated by the
analytical expressions
\begin{eqnarray}
\rho_{\rm R}(\bar{E}) &=& \exp \left( a \bar{E}^{4/7} + b \bar{E}^{1/7} \right) \\
\rho_{{\rm R}J}(\bar{E},J) &=& \rho_{\rm R}(\bar{E}) (2J+1)
\sqrt{\frac{\beta(\bar{E})^3}{\pi}}
\exp \left( -\beta(\bar{E}) (J + 1/2)^2 \right) \\
\beta(\bar{E}) &=& c \bar{E}^{-8/7} + d \bar{E}^{-13/7} 
\end{eqnarray}
if the values a = 2.5088, b = -4.3180, c = 0.8642 and d = -0.3524 are used~\cite{Note1}.
Note that for fixed $J$, the density is the number of distinct states, i.e.
does not include the $(2J+1)$ spatial quantization degeneracy of each such state.
We define the ripplon microcanonical temperature, $T_{\rm R}(\bar{E})$,
and with fixed total energy and angular momentum, $T_{{\rm R}J}(\bar{E},J)$, by
\begin{eqnarray}
\frac{1}{T_{\rm R}(\bar{E})} &=& \frac{d \ln \left( \rho_{\rm R}(\bar{E}) \right)}{d\bar{E}}
= \frac{4}{7} a \bar{E}^{-3/7} + \frac{1}{7} b \bar{E}^{-6/7}
\\
\frac{1}{T_{{\rm R}J}(\bar{E},J)} &=& \frac{d \ln \left( \rho_{{\rm R}J}(\bar{E})
\right)}{d\bar{E}} =
\frac{1}{T_{\rm R}(\bar{E})} +
\left( \frac{3}{2\beta(\bar{E})} -(J + 1/2)^2 \right) \left( \frac{d\beta}{d\bar{E}} \right)
\end{eqnarray}
where in both cases, temperature is measured in units of reduced ripplon energy, $E_{\rm
R}$, divided by Boltzmann's constant.

We now consider a droplet that has a solvated rigid linear rotor, with effective rotational
constant $B$, given again in units of $E_{\rm R}$.  Let $j$ be the rotational quantum
number of the rotor, $J_{\rm R}$ the rotational quantum number of the ripplons, and 
$J$ the total rotational quantum number.   The total density of states for fixed values
of $\bar{E}$, $J$ is given by
\begin{equation}
\rho(\bar{E},J) = \sum_j \sum_{J_{\rm R} = |J - j|}^{J + j} 
\rho_{\rm{R}J}\left(\bar{E} - Bj(j+1),J_{\rm R} \right)
\end{equation}
$P(j)$, the fraction of states with rotor quantum number $j$, is given by
\begin{equation}
P(j) = \frac{ \sum_{J_{\rm R} = |J - j|}^{J + j} \rho_{\rm{R}J}\left(\bar{E} -
Bj(j+1),J_{\rm R} \right) }{\rho(\bar{E},J)}     
\end{equation}
The temperature, $T(\bar{E},J)$, of the coupled system, at fixed $\bar{E}$ and $J$ is given
by
\begin{equation}
\frac{1}{T(\bar{E},J)} = \frac{d \ln \left( \rho(\bar{E},J) \right) }{d\bar{E}} =
\frac{ \sum_j \sum_{J_{\rm R} = |J - j|}^{J + j} 
\rho_{\rm{R}J}\left(\bar{E} - Bj(j+1),J_{\rm R} \right) \frac{1}{T_{{\rm R}J}(\bar{E})}
  }{\rho(\bar{E},J)}
\end{equation}

A rotor temperature, $T_j(\bar{E},J)$ can be defined from a Boltzmann fit
to the rotor level populations $P(j)$.  This latter corresponds to the ``droplet
temperature'' that has been reported in numerous experiments.  This assignment of
the rotor temperature to that of entire system, i.e. that $T_j(\bar{E},J) =
T(\bar{E},J)$ is natural assumption for a system in microcanonical equilibrium,
\textit{i.e.}~by equating $P(j)$ to the ratio of density of states.  If
we ignore angular momentum conservation and sum over all values of $J$,
then the above identification would be exact as long as the mean energy in the rotor,
$T_r$, if much less than the total energy in the droplet, $\bar{E}$, so that 
the ``heat bath'' of ripplon states does not change temperature 
significantly over the range of significantly populated rotor states.
Equivalently, we expect $T_j(\bar{E},J) = T(\bar{E},J)$ if that $\rho_{\rm R}(\bar{E})$ can
be approximated by $\rho_{\rm R}(\bar{E} - Bj(j+1)) = \rho_{\rm R}(\bar{E}) \exp
\left( -Bj(j+1)/T_{\rm R}(\bar{E}) \right)$.  Note that if the finite heat capacity of
the droplet heat bath is considered, then we would expect the population in higher
rotational levels to fall off faster than predicted by the Boltzmann distribution,
and thus the effective temperature determined by a fit to the rotor populations to
be lower than the droplet temperature.

When we look at the $J$ conserving ensemble, we must consider the fact that 
the average value of 
$J_{\rm R}(J_{\rm R} +1)$ equals to $J(J+1) - j(j+1)$ for fixed $J, j$.  Thus the population of 
different rotor levels will be influenced not only by $T_{\rm R}$, but also
by the $J$ dependent factor in $\rho_{{\rm R}J}$, which has the same form as the rotational
distribution function of a spherical top in a canonical ensemble.  In
order to test the size of the expected bias in the estimate of $T(\bar{E})$
by $T_j$, we will examine numerical results for droplet of $N = 3\,000$ and $10\,000$ helium
atoms, for which $E_{\rm R} = 0.069$ and $0.0377$\,K respectively.  Based upon our
evaporative cooling calculations~\cite{Dokter03}, we will assume that the
droplets cool to a final temperature of 0.35\,K, but have a wide range of final angular
momentum values, $J$.  This condition allows calculation of an isothermal curve $\bar{E}(J)$
for each droplet size.
We further assume that $B = 1\,\textrm{GHz}$, which equals 0.70 (1.27) in reduced units
for droplets of 3\,000 (10\,000) helium atoms.
This is the effective $B$ constant for SF$_6$ dissolved in helium, from which the
temperature of helium nanodroplets was first measured~\cite{Hartmann95}.  It is found that
$T(\bar{E}(J),J) = T_{{\rm R}J}(\bar{E}(J),J)$ to better than 1\,\% for all values of $J$
over the range from 0--4000.  It is found that the values of
$P(j)$ fit a Boltzmann distribution to high accuracy, so that values of $T_r$ are 
well defined.
However, over this range of $J$, we find a small be systematic difference in the values of
between $T_j(\bar{E}(J),J)$ and $T(\bar{E}(J),J)$, as demonstrated in figure 1.  
The ratio of $T_j(\bar{E}(J),J)$ to $T(\bar{E}(J),J)$ is found to be approximately
proportional to $B^{-1}$, i.e. is linearly proportional to the effective rotational
moment of inertia of the rotor in helium (including the contribution due to motion
of the helium).  As is evident from the figure, the size of the bias decreases for larger
droplets, though less than linearly with number of helium atoms in the droplet.

The present results demonstrate that for helium droplets with significant trapped
angular momentum, the population of rotational levels of rotors will be 
biased from that predicted by a Boltzmann distribution at the temperature of
either the droplet excitation modes or the temperature of the entire system.   The size of the
bias is on the order that should be detectable by careful measurement of the droplet 
rotational constant as the temperature of the pickup gas (and thus the initial 
collisional angular momentum) is increased.   Previous experiments have
reported a range of rotational temperatures, but up to know these do not appear to have
been considered physically significant.  
For example, in a study of the cluster size dependence of the spectrum of SF$_6$,
the temperature was found to be constant to within statistical noise  ($0.36-0.40$\,K)
for droplets of more than a few thousand He atoms, but to systematically increase for
smaller droplets, down to a value of 0.48(20)\,K for droplets of a 
few hundred He atoms~\cite{Hartmann99}.  The authors concluded that the 
temperature of smaller droplets was higher, but the present result suggest that
another possibility is that the effect they observed could be, at least in part,
due to the bias effect predicted by the present work.
Nauta and Miller found that for the spectrum of HF dimer, the rotational temperature
was 0.34(1)\,K~\cite{Nauta00b}, below the excepted droplet temperature of
0.38(2)\,K.  It should be pointed out that the rotational constant of HF dimer in helium is
three times larger than that of SF$_6$, and that the pickup of two HF molecules should
deposit  considerably less angular momentum than the pickup of SF$_6$.
It should be noted that the size of the bias is a
function only of the total rotational quantum number, and thus the effect should be
essentially unchanged for a symmetric top or spherical top molecule.    For an asymmetric
top, this effect should lead to a failure of the transition  intensities to follow that
expected for a gas in thermal equilibrium if this bias effect is ignored.
Stimulated by the present analysis, Roger Miller and collaborators have made an
attempt to look in a systematic way at the temperature extracted from fits to
rotor populations for different molecules.   Unfortunately, the asymmetries of
the lineshapes often observed for the lowest rovibrational transitions lead to
model dependent biases in the fitted rotational temperatures that could not be
disentangled from the expected effects~\cite{RMiller_pc}. 
This give yet one more reason for the need to address this problem of
lineshapes of ro-vibrational transitions in helium nanodroples beyond the
one previous attempt~\cite{Lehmann99b}.   That previous work assumed 
a canonical ensemble of internal states which the recent evaporative cooling
work has now shown to be a poor assumption.  

This work was supported by a grant from the National Science Foundation.  The author would
like to acknowledge the discussions with Adriaan Dokter, Roman Schmied, and Giacinto Scoles.


\newpage

\begin{figure}[htbp]
\includegraphics[width=6in]{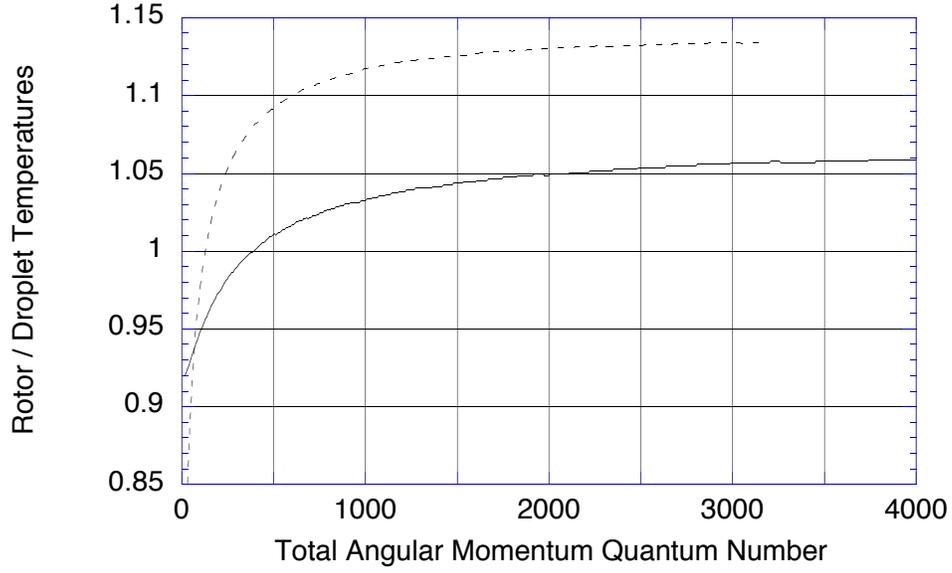}
  \caption{Plot of the ratio of the rotor temperature, $T_r(\bar{E},J)$ to
the droplet temperature, $T(\bar{E},J)$ for a droplets of 3\,000 (dashed curve)
 and 10\,000 (solid curve) helium atoms
and a solvated rotor with $B = 1$\,GHz as a function of the total angular
momentum quantum number, $J$.   For each value of $J$, $\bar{E}$ has been
fixed by the condition that $T_{{\rm R}J} = 0.35$\,K }
  \label{fig:T_ratio}
\end{figure}

\end{document}